\documentclass[12pt,aps,prb,preprint]{revtex4}

\usepackage{amsmath} 
\usepackage{graphicx} 

\usepackage{enumerate}
\newcommand{\omv}{\boldsymbol{\omega}}
\newcommand{\Omv}{\boldsymbol{\Omega}}
\newcommand{\Lv}{\boldsymbol{L}}
\newcommand{\Xv}{\boldsymbol{X}}

\newcommand{\Rv}{\boldsymbol{R}}
\newcommand{\Fv}{\boldsymbol{F}}
\newcommand{\Uv}{\boldsymbol{U}}

\begin{document}

\title{Spinning eggs---which end will rise?}
\author{Ken Sasaki}
\email{sasaki@phys.ynu.ac.jp}
\affiliation{Department of Physics, Faculty of Engineering, Yokohama National
University, Yokohama 240-8501, Japan}

\begin{abstract}
We examine the spinning behavior of egg-shaped axisymmetric bodies whose
cross sections are described by several oval curves similar to real eggs
with thin and fat ends. We use the gyroscopic balance condition of Moffatt
and Shimomura and analyze the slip velocity of the bodies at the point of
contact as a function of $\theta$, the angle between the axis of symmetry
and the vertical axis, and find the existence of the critical angle
$\theta_c$. When the bodies are spun with an initial angle
$\theta_{\rm initial}>\theta_c$, $\theta$ will increase to $\pi$, implying
that the body will spin at the thin end. Alternatively, if
$\theta_{\rm initial}<\theta_c$, then
$\theta$ will decrease. For some oval curves, $\theta$ will reduce to 0 and
the corresponding bodies will spin at the fat
end. For other oval curves, a fixed point at $\theta_f$ is predicted, where
$0 <\theta_f< \theta_c$. Then the bodies will spin not at the fat end, but
at a new stable point with
$\theta_f$. The empirical fact that eggs more often spin at the fat
than at the thin end is explained.
\end{abstract}

\maketitle

\section{Introduction}
Spinning objects have historically been interesting subjects to study. The
spin reversal of the rattleback\cite{GH} (also called a celt or
wobblestone) and the behavior of the tippe top\cite{GN} are typical
examples. Recently, the riddle of spinning eggs was resolved by Moffatt and
Shimomura.\cite{MS} When a hard-boiled egg is spun sufficiently rapidly on
a table with its axis of symmetry horizontal, the axis will rise from the
horizontal to the vertical. They discovered that if an axisymmetric body is
spun sufficiently rapidly, a gyroscopic balance condition holds. Given this
condition a constant of the motion exists for the spinning motion of an 
axisymmetric body. The constant, which is known as the Jellett
constant,\cite{Jellett} has been found previously for symmetric tops such
as the tippe top. Using these facts, they derived a first-order
differential equation for $\theta$, the angle between the axis of symmetry
and the vertical axis. For a uniform spheroid as an example they showed
that the axis of symmetry indeed rises from the horizontal to the vertical.

The shape of an egg looks like a spheroid, but is not exactly so. It has thin and
fat ends. Which end of the spinning egg will rise? Empirically, we know that
either end can rise. But we more often see eggs spinning at the fat end with
the thin end up rather than the other way round. In this paper we
investigate the spinning behavior of egg-shaped axisymmetric bodies whose
cross sections are described by several oval curves. We use the gyroscopic
balance condition and analyze the slip velocity of the body at the point of
contact as a function of
$\theta$ and find the existence of the critical angle $\theta_c$ for each
model curve. When the bodies are spun with the initial angle $\theta_{\rm
initial}>\theta_c$,
$\theta$ will increase to $\pi$, which means that the body will spin at the
thin end. Alternatively, if
$\theta_{\rm initial}<\theta_c$, then $\theta$ will decrease. For some oval
curves, $\theta$ will decrease to 0 and the corresponding bodies will shift
to the stable spinning state at the fat end. For other oval curves, a fixed
point at 
$\theta_f$ is predicted, where $0 <\theta_f< \theta_c$. In this case the
bodies will spin not at the fat end but at a new stable point with
$\theta_f$. We also explain why we observe more eggs spinning at the fat
end than at the thin end.

The paper is organized as follows: To explain our notation and the
geometry, we review the work of Ref.~\onlinecite{MS} on spinning eggs in
Sec.~II. Then in Sec.~III we introduce several models of oval curves that
we will study. In Sec.~IV we analyze the spinning behavior of
axisymmetric bodies whose cross sections are described by these oval
curves. The final section is devoted to a summary and discussion.

\section{Spinning egg}
\begin{figure}[h]
\begin{center}
\includegraphics{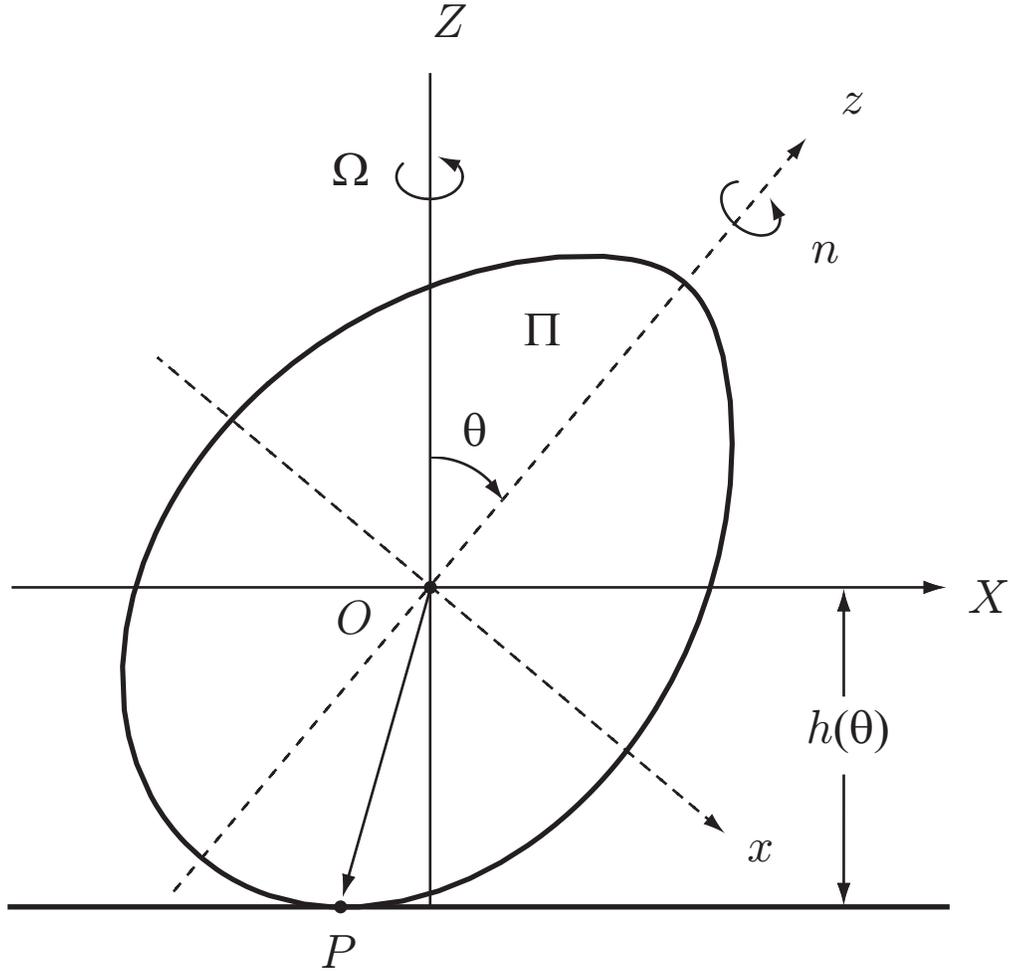}
\caption{\label{AxisymmetricBody}An axisymmetric body with center of mass
$O$ on a horizontal table with point of contact $P$. Its axis of
symmetry, $O\!z$, and the vertical axis, $O\!Z$, define a plane $\Pi$,
which precesses about $O\!Z$ with angular velocity 
$\Omv(t)=(0,0,\Omega)$. $O\!X\!Y\!Z$ is a rotating frame of
reference with $O\!X$ horizontal in the plane $\Pi$. The height of $O$ above
the table is $h(\theta)$ (from Ref.~3).}
\end{center}
\end{figure}

We follow the geometry and notation of Ref.~\onlinecite{MS} in their
analysis of spinning eggs as much as possible. As is shown in
Fig.~\ref{AxisymmetricBody}, an axisymmetric body spins on a horizontal table
with point of contact $P$. We will work in a rotating frame of reference
$O\!X\!Y\!Z$, where the center of mass is at the origin, $O$. The symmetry axis
of the body, $O\!z$, and the vertical axis, $O\!Z$, define a plane $\Pi$, which
precesses about $O\!Z$ with angular velocity 
$\Omv(t)=(0,0,\Omega)$. We choose the horizontal axis $O\!X$
in the plane
$\Pi$ and thus $O\!Y$ is vertical to $\Pi$ and inward. The angle of interest is
$\theta(t)$, the angle between $O\!Z$ and $O\!z$. 

In a rotating frame of reference $O\!xyz$, where $O\!x$ is in the plane $\Pi$ and
perpendicular to the symmetry axis $O\!z$ and where $O\!y$ coincides with
$O\!Y$, the body spins about $O\!z$ with the rate $\dot \psi$. 
Since, in the frame $O\!xyz$,  $\Omv$ is expressed as 
$\Omv= -\Omega \sin\theta \hat x +\Omega \cos\theta \hat z $, 
the angular velocity of the body, 
$\omv$, is given by $\omv= -\Omega 
\sin\theta \hat x+\dot \theta \hat y +n \hat z $.  Here ${\hat x}$, ${\hat
y}$, and ${\hat z}$ are unit vectors along $O\!x$, $O\!y$, and $O\!z$,
respectively,  $n(t)$ is
 given by $n=\Omega \cos\theta+\dot \psi$, and 
the dot represents differentiation with respect to time. The
$O\!x$ and
$O\!y$ axes are not body-fixed axes but they are principal axes,
so that the angular momentum, $\Lv$, is expressed by 
$\Lv=-A\Omega \sin \theta \hat x +A \dot \theta \hat y +Cn
\hat z$, where $(A, A, C)$ are the principal moments of inertia at $O$. 

The coordinate system $O\!X\!Y\!Z$ is obtained from the frame $O\!xyz$ by rotating the latter 
about the $O\!y$ ($O\!Y$) axis through the angle $\theta$. Hence, in the rotating frame
$O\!X\!Y\!Z$, $\omv$, and $\Lv$ have components
\begin{eqnarray}
\omv &=&\big( (n-\Omega \cos \theta)
\sin \theta, \dot \theta,\Omega \sin^2\theta +n \cos \theta
\big), \\
\Lv &=&\big( (Cn-A\Omega \cos \theta) \sin \theta, A \dot
\theta,A\Omega \sin^2\theta +Cn \cos\theta \big), \label{angular}
\end{eqnarray}
respectively. The ½ evolution of $\Lv$ is governed by Euler's
equation,
\begin{equation}
\frac{\partial \Lv}{\partial t}+\Omv\times 
\Lv =\Xv \!_P \times
(\Rv+\Fv), \label{Euler}
\end{equation} where $\Xv\!_P$ is the position vector of the contact point $P$
from $O$, 
$\Rv$ is the normal reaction at $P$, $\Rv=(0,0,R)$, with $R$
being of order $M\!g$, the weight, and 
$\Fv$ is the frictional force at $P$. Because the point $P$ lies
in the plane
$\Pi$, $\Xv\!_P$ has components $(X\!_P, 0, Z_p)$, which are given by
\begin{subequations}
\label{height}
\begin{eqnarray}
Z_P&=&-h(\theta) \\
X\!_P&=&\frac{dh}{d\theta}, \label{height.b}
\end{eqnarray}
\end{subequations}
where $h(\theta)$ is the height of $O$ above the table. We will see in
Sec.~IV that
$h(\theta)$ is determined as a function of 
$\theta$, once the geometry and density distribution of the body are known. 

When the frictional force is weak and $\dot \theta$ is correspondingly small,
the slip velocity of the point $P$ is, to leading order in $\dot \theta$,
expressed as 
$\Uv \!_P=(0,V_P,0)$, where
\begin{equation}
V_P= (\Omega \sin^2\theta+n \cos\theta)\frac{dh}{d\theta}+ 
(n-\Omega \cos\theta)h(\theta) \sin \theta. \label{Vpfirst}
\end{equation}
Hence, the frictional force, $\Fv$, is to leading order, 
$\Fv=(0,F,0)$, where $F$ is a function of $V_P$ given by the law of dynamic
friction between the two surfaces in contact. We assume later Coulomb 
friction for $F$.

The $Y$-component of Eq.~(\ref{Euler}) is expressed by 
\begin{equation}
A{\ddot \theta}+(Cn-A\Omega \cos\theta)\Omega \sin\theta
 =-RX\!_P~.
\label{Ycomponent}
\end{equation}
Since the secular change of $\theta$ with which we are concerned is slow and
thus $|\ddot \theta|\ll \Omega^2$, the first term of  Eq.~(\ref{Ycomponent})
can be neglected. Furthermore, in a situation where $\Omega^2$ is
sufficiently large so that the terms involving $\Omega$ in
Eq.~(\ref{Ycomponent}) dominate the term $-RX\!_P$, Eq.~(\ref{Ycomponent}) 
is reduced, at leading order,  to $(Cn-A\Omega \cos\theta)\Omega
\sin\theta=0$. Then, for $\sin \theta\not= 0$, we arrive at a condition 
\begin{equation}
Cn=A\Omega \cos\theta,
\label{Condition}
\end{equation}
which was recently discovered by Moffatt and Shimomura\cite{MS} and was called
by them as the gyroscopic balance condition. 
Under this condition, the Jellett constant\cite{Jellett} 
also exists for a general axisymmetric body.  With Eq.~(\ref{Condition}), the
angular momentum simplifies to
$\Lv=( 0, A{\dot \theta}, A\Omega)$, and the $X$- and
$Z$-components of Eq.~(\ref{Euler}) reduce, respectively, to
\begin{subequations}
\label{Euler2}
\begin{eqnarray}
A\Omega\dot\theta=FZ\!_P, \label{Euler2.a} \\
A\dot\Omega=FX\!_P.
\end{eqnarray}
\end{subequations}
Equation~(\ref{Euler2}), together with Eq.~(\ref{height}), leads to 
\begin{equation}
-\Lv \cdot \Xv \!_P=A\Omega h=J = \mbox{constant}, \label{Jellett}
\end{equation}
where the Jellett constant $J$ is determined by the initial conditions. From
Eqs.~(\ref{height}), (\ref{Jellett}), and (\ref{Euler2.a}),
we obtain a first-order differential equation for $\theta$, 
\begin{equation} J\dot\theta=-Fh^2(\theta). \label{EquForTheta}
\end{equation} If we assume Coulomb friction, $F$ is given by
\begin{equation} F=-\mu Mg\frac{V_P}{\vert V_P \vert}, \label{Friction}
\end{equation} and $V_P$ in Eq.~(\ref{Vpfirst}), given the gyroscopic
balance condition (\ref{Condition}), is expressed as a function of $\theta$
as: 
\begin{equation}
V_P=\frac{J}{Ah(\theta)}\bigg [\Big( \sin^2\theta +\frac{A}{C}
\cos^2\theta \Big)\frac{dh}{d\theta}+ \sin \theta \cos\theta
\Big(\frac{A}{C}-1 \Big)h(\theta)\bigg ]. \label{Vpsecond}
\end{equation}
Hence, if we know $h(\theta)$ from geometrical considerations , we may
solve Eq.~(\ref{EquForTheta}) and determine the time-dependence of
$\theta$. Moffatt and Shimomura\cite{MS} considered a uniform spheroid as
an example and showed that $\theta$ decreases from $\pi/2$ to 0 for the
prolate spheroid while $\theta$ increases from 0 to $\pi/2$ for the oblate
one.

However, the shape of an egg is not a spheroid and has thin and fat ends.
Which end of the spinning egg will rise? Empirically, we know that either
end may rise and that the body spins with its axis of symmetry vertical.
Which end the spinning egg chooses might seem to depend on the inclination
of the initial axis of symmetry, that is, the initial value of
$\theta$. In the following we examine several models of oval curves and
determine the relation between the initial value of
$\theta$ and the final spinning position of the egg-shaped body.

\section{Models of oval curve}
The shape of a three-dimensional egg can be reconstructed by rotating
its two-dimensional cross section around the axis of symmetry. The cross
section of an egg looks similar to an ellipse, but is not quite. It is
sharper at one end than at the other. We will examine several model
curves that have been proposed for the cross-section of a real egg.

Let us consider an axisymmetric body whose cross-section is described by
\begin{equation}
\label{CrossSection}
x^2=g(z),
\end{equation}
with $g(z)> 0$ for $z_{\min} < z
< z_{\max}$ and $g(z_{\min})=g(z_{\max})=0$,
where we choose the $z$ axis as the symmetry axis. If the body
has uniform density, then the volume and the $z$ component of center of
mass are given, respectively, by
\begin{subequations}
\begin{eqnarray}
V&=&\pi\!\int^{z_{\max}}_{z_{\min}}g(z)\,dz, \\
z_g&=&\frac{\pi}{V}\!\int^{z_{\max}}_{z_{\min}}zg(z)\,dz.
\end{eqnarray}
\end{subequations}
The principal moments of inertia at center of mass are expressed by
\begin{subequations}
\begin{eqnarray}
A&=&M\frac{\pi}{V} \! \int^{z_{\max}}_{z_{\min}}
\bigg [\frac{1}{4}[g(z)]^2+g(z)(z-z_g)^2 \bigg ]dz, \\
C&=&\frac{M}{2}\frac{\pi}{V}\!\int^{z_{\max}}_{z_{\min}}[g(z)]^2\,dz.
\end{eqnarray}
\end{subequations}
Of course, the density of a real egg is not uniform. But if the density 
distribution is given by $\rho(r,z)$ as a function of $r$ and $z$, where
$r$ is the distance from the symmetry axis, and the cross-section is still
described by Eq.~(\ref{CrossSection}), we can calculate $z_g$, $A$, and $C$.

The following are the oval curves that we will examine.

\begin{figure}[h]
\begin{center}
\includegraphics{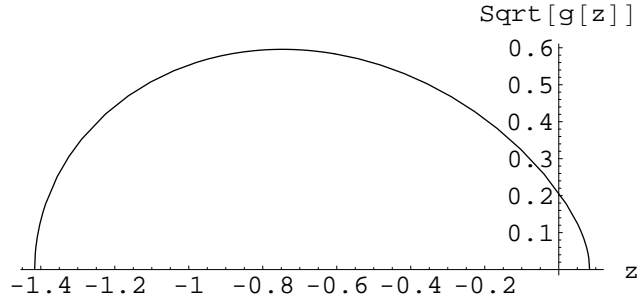}
\caption{\label{CartesianOval} A Cartesian oval in units of the arbitrary
length $a$.}
\end{center}
\end{figure}

\begin{figure}[h]
\begin{center}
\includegraphics{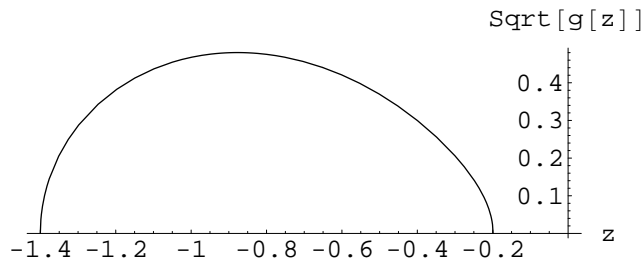}
\caption{\label{CassiniOval} A Cassini oval in units of $a$.}
\end{center}
\end{figure}

\begin{figure}[h]
\begin{center}
\includegraphics{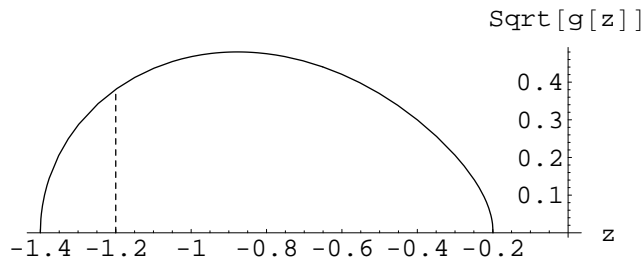}
\caption{\label{CassiniOvalAir} A Cassini oval with air chamber in units
of $a$.}
\end{center}
\end{figure}

\begin{figure}[h]
\begin{center}
\includegraphics{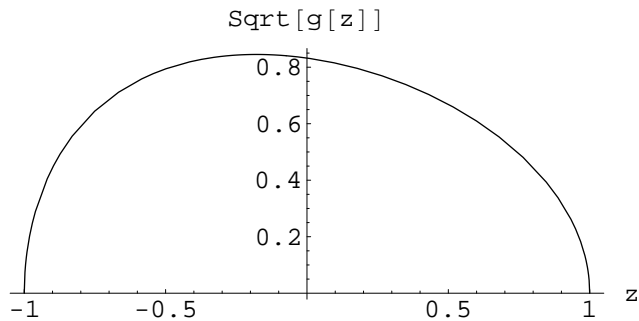}
\caption{\label{EggCurve3} Wassenaar egg curve in units of $a$.}
\end{center}
\end{figure}

\begin{figure}[h]
\begin{center}
\includegraphics{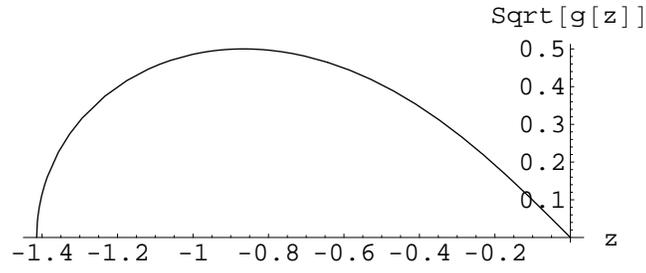}
\caption{\label{Lemniscate} The lemniscate in units of $a$.}
\end{center}
\end{figure}

\begin{enumerate}[(i)]

\item {\it Cartesian oval}. The curve, given by
$\sqrt{z^2+x^2}+m\sqrt{(z+a)^2+x^2}=c$, consists of two ovals. For definiteness we set $m=2$.
The inside oval is expressed by $x^2=g(z)$ with
\begin{equation} g(z)=- z^2-\frac{8}{3}az+\frac{a^2}{9}(5\kappa^2-12)
-\frac{4}{9}\kappa a^2 \sqrt{\kappa^2-3-\frac{6z}{a}},
\label{EqCartesianOval}
\end{equation}
with
$\kappa=c/a$. For $\kappa=9/4$, we find that 
$g(z)$ is defined for the interval $-\frac{17}{12}a\le z \le \frac{1}{12}a$,
$z_g=-0.710a$, and $A/C=1.26$ (see Fig.~\ref{CartesianOval}).

\item {\it Cassini oval}. This quartic curve is expressed by 
$[(z+a)^2+x^2 ] [(z-a)^2+x^2 ]=b^4$ with $a, b > 0$. If
$a>b$, the curve consists of two loops, both of which look like the
cross-section of a real egg with thin and fat ends. We choose the one that
is expressed by $x^2=g(z)$ with
\begin{equation}
\label{EqCassiniOval}
g(z)=-(z^2+a^2)+a\sqrt{4z^2+\lambda^4 a^2},
\end{equation}
where $\lambda=\frac{b}{a} <1 $ for
$-a\sqrt{1+\lambda^2}\le z \le -a\sqrt{1-\lambda^2}$,
so that the thin end points to the positive $z$ axis (see
Fig.~\ref{CassiniOval}). For
$\lambda=0.98$, we find $z_{\min}=-1.40a$, $z_{\max}=-0.199a$, $z_g=-0.840a$, and
$A/C=1.25$.

\item {\it Cassini oval with an air chamber}. A real egg has an air
chamber near the fat end. We take into account the existence of an air
chamber by using the Cassini oval (\ref{EqCassiniOval}) and taking
$z_{\min}=-\alpha a$, with $\sqrt{1-\lambda^2}<\alpha<\sqrt{1+\lambda^2}$
for the evaluation of $V$,
$z_g$, $A$, and $C$. This condition means that an empty space exists for
$-\sqrt{1+\lambda^2}a\le z\le -\alpha a$ (see Fig.~\ref{CassiniOvalAir}).
 For 
$\lambda=0.98$ and $\alpha=1.2$, we obtain $z_g=-0.798a$ and $A/C=1.07$. The
position of the center of mass $z_g$ is closer to the thin end and the ratio
$A/C$ is smaller compared to the curve without an air chamber.

\item {\it Wassenaar egg curve}. A rather simple equation for an oval curve
was proposed recently by Wassenaar\cite{Wassenaar} and is given by
\begin{equation}
x^2=g(z)=2a\big [-2z-\xi a+\sqrt{4a^2+4\xi az+\xi^2 a^2}
\big ],
\label{EqWassenaar}
\end{equation}
for $5<\xi <6 $ and $-a\le z \le a$.
For $\xi=5.6$ we find $z_g=-0.0714a$ and $A/C=1.21$ (see
Fig.~\ref{EggCurve3}).

\item {\it Lemniscate of Bernoulli}. The lemniscate of Bernoulli is not
a candidate for oval curves and actually looks like the infinity symbol. We
study it because its final spinning position might be
interesting. The curve is expressed by 
$[(z+a)^2+x^2 ] [ (z-a)^2+x^2]=a^4$. We study the half of the curve that
is given by 
\begin{equation}
x^2=g(z)=-(z^2+a^2)+a\sqrt{4z^2+a^2},
\label{EqLemniscate}
\end{equation}
with $ -\sqrt{2}a\le z
\le 0$. The lemniscate is a special case of a Cassini oval and is obtained
by setting $a=b$ in Eq.~(\ref{EqCassiniOval}). For the lemniscate, we obtain
$z_g=-0.813a$ and
$A/C=1.34$ (see Fig.~\ref{Lemniscate}).

\end{enumerate}

\begin{figure}[h]
\begin{center}
\includegraphics{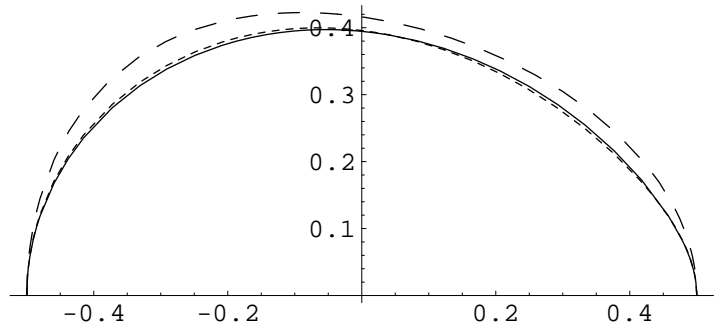}
\caption{\label{All} The Cartesian oval (a solid curve), Cassini oval
(short-dashed curve), and a proposed egg curve (dashed curve) adjusted to
have the same length along the symmetry axis.}
\end{center}
\end{figure}

We plot in Fig.~\ref{All} the 
Cartesian and Cassini ovals and the Wassenaar egg curve, adjusting the
parameter $a$ for each case so that they have the same length along the
symmetry axis. We observe that the Cartesian and Cassini ovals almost
overlap. Indeed the axisymmetric bodies whose cross sections are expressed
by these oval curves have close values of
$A/C$ (1.26 and 1.25 for the Cartesian and Cassini ovals, respectively).
However, we will see in Sec.~\ref{sec4} that these ovals predict different
spinning behavior for the corresponding axisymmetric bodies.

\section{Which end will rise?}\label{sec4}
We obtain from Eqs.~(\ref{EquForTheta}) and (\ref{Friction}) 
\begin{equation}
\label{eq:theta}
\dot\theta=\frac{\tau }{\vert V_P \vert} {\widetilde V}_P, 
\end{equation}
with
\begin{equation}
{\widetilde V}_P=V_P\frac{A}{J} \quad {\rm and} \quad \tau=\frac{h^2\mu
Mg}{A}.
\end{equation}
Equation~(\ref{eq:theta}) implies that the change of $\theta$ is governed by
the sign of
${\widetilde V}_P$. If
${\widetilde V}_P$ is positive (negative), $\theta$ will increase
(decrease) with time. Therefore a close examination of the behavior of
${\widetilde V}_P$ as a function of $\theta$ will be important. Moffatt
and Shimomura\cite{MS} showed that for a uniform prolate spheroid,
${\widetilde V}_P$ has the form ${\widetilde V}_P\propto 
\sin 2\theta$ with a negative proportionality constant. Thus if the body is
spun (sufficiently rapidly) on a table with the initial inclination angle
$\theta_{\rm initial} < \pi/2$, then $\theta$ decreases to 0. On the other
hand, if the body is spun with
$\theta_{\rm initial}>\pi/2$, $\theta$ increases to $\pi$. Either
end will rise, because both ends of the prolate spheroid look the same. The
case of a real egg is different. We can easily distinguish between the thin
and the fat end. We will now analyze the axisymmetric bodies whose
cross-sections are expressed by the oval curves introduced in Sec.~III.

\begin{figure}[h]
\begin{center}
\includegraphics{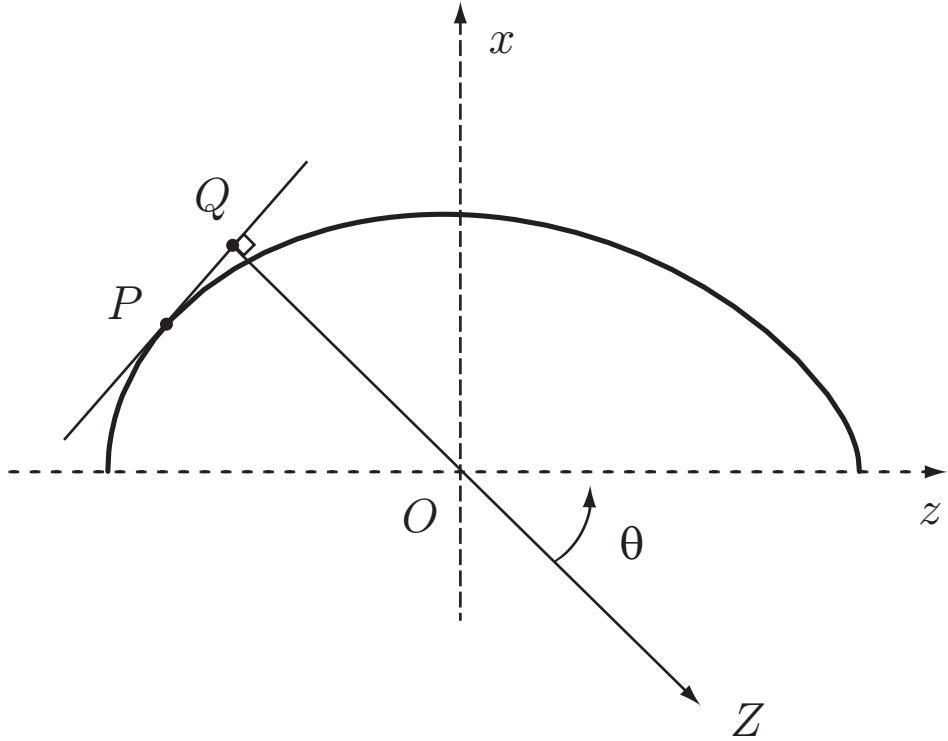}
\caption{\label{Egg_b} An oval curve $x=\sqrt{f(z)}$ that describes a
part of the cross section of an axisymmetric body. The center of mass $O$
is at the origin. The line
$OQ$ is perpendicular to the line tangent to the curve at the point $P(z,
x=\sqrt{f(z)})$. The line $PQ$ is on a horizontal table and
$P$ corresponds to the point of contact. The line $QO$ defines the
vertical axis $OZ$, and the polar angle $\theta$ between
$OZ$ and $Oz$ is given by $\tan \theta =1/\beta$, where $\beta$ is the slope of
the line tangent to the curve at $P$. The length of $OQ$ is $h(\theta)$.}
\end{center}
\end{figure}

We take a coordinate system in which the center of mass $O$ resides at the origin. In this
coordinate system, the oval curves satisfy 
\begin{equation} x^2=f(z)=g(z+z_g),
\end{equation} where $g(z)$ is introduced in Eq.~(\ref{CrossSection}) to
describe the cross-section of an axisymmetric body whose center of mass is
at $z=z_g$. We consider the point $P$($z,x=\!\sqrt{f(z)} $) on the curve
(see Fig.~\ref{Egg_b}). The slope
$\beta$ of the line tangent to the curve at $P$ is given by
\begin{equation}
\beta\equiv \frac{dx}{dz} =\frac{f'(z)}{2\sqrt{f(z)}}. 
\end{equation} Draw a line from the origin which is perpendicular to the line tangent to the 
curve $x=\sqrt{f(z)}$ at $P$. Let the point of intersection be $Q$($z_Q, x_Q$), whose
coordinates are
\begin{subequations}
\begin{eqnarray}
z_Q&=&\frac{\beta}{\beta^2+1}(\beta z-\sqrt{f(z)}),\\
x_Q&=&-\frac{1}{\beta}z_Q.
\end{eqnarray}
\end{subequations}
Suppose that the line $PQ$ is in a horizontal plane of table and 
$P$ is the point of contact. Then the line $QO$ defines the vertical axis $OZ$. The polar
angle, $\theta$, between $OZ$ and $Oz$ is determined by,
\begin{equation}
\tan \theta
=\frac{1}{\beta}=\frac{2\sqrt{f(z)}}{f'(z)},\label{Theta}
\end{equation} which gives the relation between $\theta$ and $z$. The height, $h(\theta)$, of
$O$ above the table is equal to the length of $OQ$, and we obtain 
\begin{equation}
h(\theta)=\sqrt{z_Q^2+x_Q^2}=\frac{1}{\sqrt{\beta^2+1}}(\sqrt{f(z)}-\beta
z),
\end{equation} because $(\sqrt{f(z)}-\beta z)>0$. The squared length of $PQ$
corresponds to
$X\!_P^2$. We choose the sign of $X\!_P$ to be the same as that of
$(z-z_Q)$ and obtain
\begin{equation}
X\!_P=\frac{1}{\sqrt{\beta^2+1}}\Big(z+\beta\sqrt{f(z)}\Big).\label{XP}
\end{equation}
If we use Eqs.~(\ref{Theta})--(\ref{XP}), we confirm
Eq.~(\ref{height.b}) and find
that $V_P$ in Eq.~(\ref{Vpsecond}) can be rewritten as a function $z$ as
follows:
\begin{equation}
V_P
=\frac{J}{A}\frac{\beta^2}{\beta^2+1} \bigg[\Big(\frac{1}{\beta^2}+\frac{A}{C} 
\Big) \frac{z+\beta \sqrt{f(z)}}{\sqrt{f(z)}-\beta z} +
\frac{1}{\beta}\Big(\frac{A}{C}-1 \Big) \bigg].\label{Vpthird}
\end{equation}

\begin{figure}[h]
\begin{center}
\includegraphics{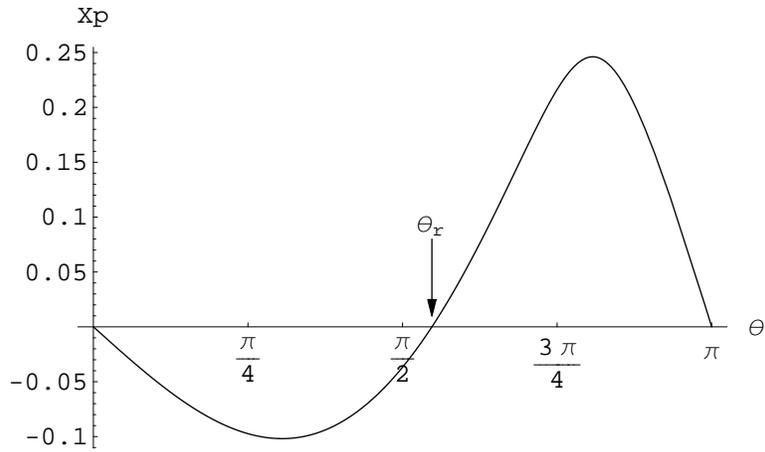}
\caption{\label{CartesianOvalXpTheta} $X_P$ as a function of $\theta$ for a Cartesian oval.}
\end{center}
\end{figure}

{}From Eq.~(\ref{Theta}), $X_P$ and $V_P$ can be considered as
functions of $\theta$. As an example, we plot in
Fig.~{\ref{CartesianOvalXpTheta}} the graph of $X_P$ versus 
$\theta$ for the case of the Cartesian oval in Eq.~(\ref{EqCartesianOval}).
In addition to 
$\theta=0$ and
$\pi$,
$X_P$ vanishes at an angle $\theta_r$, which is obtained by solving 
\begin{equation} z+\beta\sqrt{f(z)}=0. \label{XPzero}
\end{equation}
When the body is placed at rest on a table, its inclination angle is
$\theta_r$ and the height 
$h(\theta)$ of center of mass $O$ from the table is a minimum at
$\theta_r$. We observe from
Eqs.~(\ref{Vpfirst}) or (\ref{Vpsecond}) that $V_P=0$ at $\theta=0$ and
$\pi$, because
$\sin\theta=0 $ and $\frac{dh}{d\theta}(=X_p)=0$ at these points. Moreover,
$V_P$ vanishes at other angles, which are given by solving
\begin{equation}
\frac{A}{C}+\frac{z}{\beta \sqrt{f(z)}}=0. \label{VPzero}
\end{equation} When $A=C$, Eqs.~(\ref{XPzero}) and (\ref{VPzero}) become
equivalent, which means that $V_P$ and $X_P$ vanish at the same inclination
angle.

We next examine the graph of ${\widetilde V}_P$ as a function of $\theta$
for the oval curves introduced in Sec.~III. 

\begin{figure}[h]
\begin{center}
\includegraphics{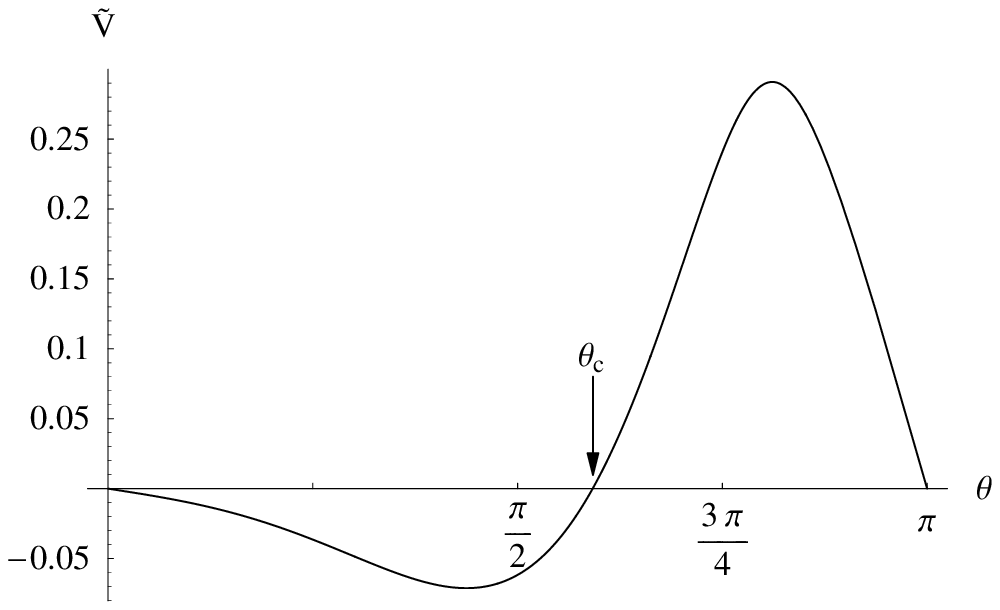}
\caption{\label{CartesianOvalVpTheta} ${\widetilde V}_P$ as a function of
$\theta$ for a Cartesian oval.}
\end{center}
\end{figure}

\medskip (i) {A Cartesian oval}. Figure~\ref{CartesianOvalVpTheta} shows
that ${\widetilde V}_P$ crosses the line
${\widetilde V}_P=0$ at an angle $\theta_c$ and 
${\widetilde V}_P>0$ for $\theta_c< \theta <\pi$ but is negative
for
$0< \theta <\theta_c$. So, the angle $\theta_c$ is a {\it critical point}.
If the body is spun on a table with the initial angle $\theta_{\rm
initial}>\theta_c$, then $\theta$ will increase to
$\pi$, which means that the body will eventually spin at the thin end.
For
$\theta_{\rm initial}<\theta_c$, we will see that the body spins at the
fat end. That is, depending on the initial value 
$\theta_{\rm initial}$, the body will spin at the thin or the fat end.
Both ends are stable points. We have $A>C$ for Cartesian ovals, which leads
to $\theta_c>\theta_r$. Numerically we obtain from Eqs.~(\ref{VPzero}) and
(\ref{XPzero}) that $\theta_c=1.86$ and
$\theta_r=1.72$. Recall that $\theta_r$ is the inclination angle when the body is placed at
rest. If we spin the body involuntarily, the initial angle
$\theta_{\rm initial}$ tends to be near 
$\theta_r$. Because $0<\theta_r<\theta_c$, it is likely that the body is spun
with
$\theta_{\rm initial}$ between 0 and $\theta_c$, and thus it will shift to
the stable spinning state at the fat end. Empirically, we more
often observe eggs spinning at the fat end rather than at the thin end. The
expected behavior of the axisymmetric body expressed by the Cartesian oval
in Eq.~(\ref{EqCartesianOval}) well explains the observed features of the
spinning egg.

\begin{figure}[h]
\begin{center}
\includegraphics{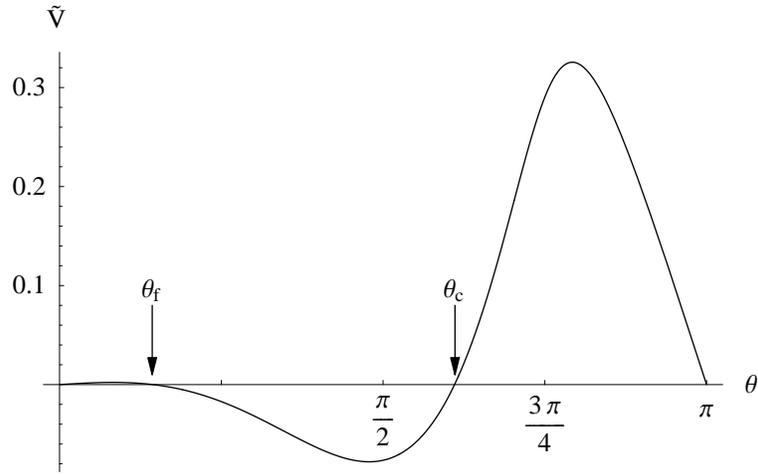}
\caption{\label{CassiniOvalVpTheta} ${\widetilde V}_P$ as a function of
$\theta$ for a Cassini oval.}
\end{center}
\end{figure}

\medskip (ii) A Cassini oval. The second example of oval curves
presents an interesting situation. We see from
Fig.~\ref{CassiniOvalVpTheta} that ${\widetilde V}_P$ crosses
the line 
${\widetilde V}_P=0$ at $\theta_f$ and $\theta_c$ ($\theta_f
<\theta_c$), and that ${\widetilde V}_P$ is negative for 
$\theta_f <\theta <\theta_c$ and otherwise positive. Numerically we obtain
$\theta_f=0.45$ and
$\theta_c=1.92$ (and $\theta_r=1.77$). Thus the graph of ${\widetilde V}_P$ in
Fig.~\ref{CassiniOvalVpTheta} implies that for the Cassini oval
(\ref{EqCassiniOval}) the thin end ($\theta=\pi$) is a stable point, but the
fat end is not. When the body is spun with the initial value of
$\theta$ anywhere between 0 and
$\theta_c$, $\theta$ will approach the {\it fixed point} $\theta_f$. In
other words, the body will spin not at the fat end, but at the point
with the inclination angle $\theta_f$. It is interesting to note that the
curves of the Cartesian (\ref{EqCartesianOval}) and Cassini
(\ref{EqCassiniOval}) ovals almost overlap each other when they are
adjusted to have the same length along the symmetry axis (see
Fig.~\ref{All}). But they predict different behaviors for ${\widetilde
V}_P$ and thus different spinning behaviors for the corresponding
axisymmetric bodies. Does a hard boiled egg show the behavior predicted by
this Cassini oval? We are almost certain that we have never seen such a
behavior.

\begin{figure}[h]
\begin{center}
\includegraphics{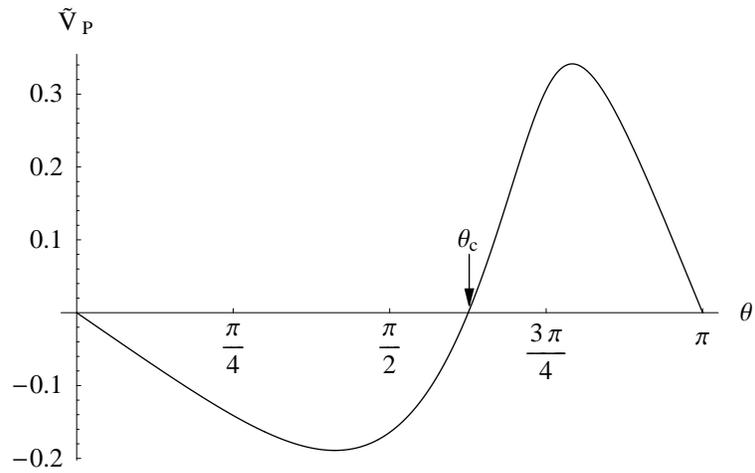}
\caption{\label{CassiniOvalAirVpTheta}${\widetilde V}_P$ as a function of
$\theta$ for a Cassini oval with an air chamber.}
\end{center}
\end{figure}

\medskip (iii) A Cassini oval with an air chamber. Because an
egg has an air chamber near the fat end, we study the case of a
Cassini oval with an air chamber. The existence of an air chamber moves the
position of center of mass $z_g$ toward the thin end, from $z_g=-0.840a$ to
$z_g=-0.798a$, and reduces the ratio $A/C$ from 1.25 to 1.07. 
Consequently, the fixed point at
$\theta_f$, which is present for 
Eq.~(\ref{EqCassiniOval}), disappears. The graph of ${\widetilde V}_P$ in
Fig.~\ref{CassiniOvalAirVpTheta} shows that it crosses the line
${\widetilde V}_P=0$ only once at
$\theta_c=1.97$. The inclination angle at rest becomes $\theta_r=1.93$,
which is very close to the value of $\theta_c$ because $A/C\approx 1$. The
axisymmetric body described by a Cassini oval with an air chamber also
reproduces the features of the spinning egg.

\begin{figure}[h]
\begin{center}
\includegraphics{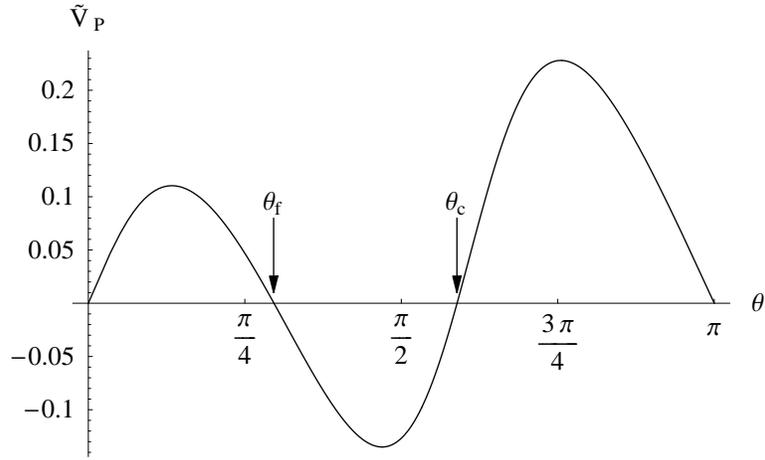}
\caption{\label{EggCurve3VpTheta} ${\widetilde V}_P$ as a function of
$\theta$ for Wassenaar egg curve.}
\end{center}
\end{figure}

\medskip (iv) Wassenaar egg curve. Figure~\ref{EggCurve3VpTheta}
shows that this curve has a fixed point at
$\theta_f=0.93$. When $\theta_{\rm initial}$ is between 0 and
$\theta_c=1.85$, 
$\theta$ will move to $\theta_f$, another stable point in addition to the
one at the thin end ($\theta=\pi$). In addition, the graph of $X_P$ for
this oval curve vanishes at two other points. One is at $\theta_r=1.78$, a
position at rest, and one at $\theta=0.58$, an unstable point.

\begin{figure}[h]
\begin{center}
\includegraphics{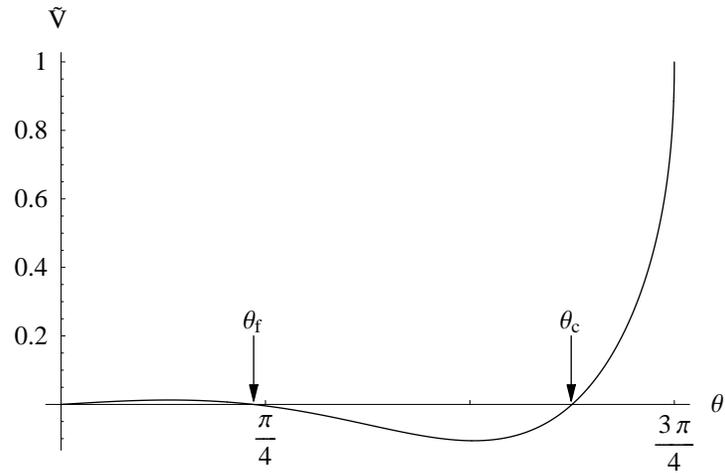}
\caption{\label{LemniscateVpTheta} ${\widetilde V}_P$ as a function of
$\theta$ for the lemniscate.}
\end{center}
\end{figure}

\medskip \noindent (v) Lemniscate of Bernoulli. For the lemniscate of
Bernoulli, we find $\lim_{z\rightarrow z_{\max}} (1/\beta) =-1$, so that
Eq.~(\ref{Theta}) tells us that the allowed region of
$\theta$ is between 0 and 
$3\pi/4$. Figure~\ref{LemniscateVpTheta} shows that
${\widetilde V}_P$ vanishes at the fat end ($\theta=0$), at the fixed point
($\theta_f=0.74$), and at the critical point ($\theta_c=1.96$). The position
of the body at rest is at
$\theta_r=1.81$, and its spinning state has two stable points at $\theta
=\theta_f$ and
$3\pi/4$. 

\section{Summary and Discussion}
We have examined the spinning behavior of axisymmetric bodies whose cross
sections are described by several model curves, including a
Cartesian oval, Cassini ovals with and without an air chamber, and
the Wassenaar egg curve. These results together
with the lemniscate of Bernoulli are summarized in Table~\ref{Summary}.
For each oval curve we used the gyroscopic balance condition
(\ref{Condition}) and found the predicted slip velocity
$V_P$ of the contact point
as a function of the inclination angle $\theta$ and the
existence of the critical angle $\theta_c$. When the body is spun on a
table with the initial angle $\theta_{\rm initial}>\theta_c$,
$\theta$ will increase to $\pi$, which means that the body will spin at
the thin end. If $\theta_{\rm initial}<\theta_c$, then
$\theta$ will decrease. For the Cartesian oval and Cassini oval with an
air chamber, $\theta$ will reduce to 0 and the corresponding bodies will 
spin at the fat end. Moreover, when the bodies are spun without intention,
we expect to see their spinning states at the fat end more often than at
the thin end because the inclination angle
$\theta_r$ at rest is smaller than $\theta_c$. This behavior is
consistent with the features of a spinning egg.

On the other hand, the Cassini oval and Wassenaar egg curves predict the
existence of the fixed point at $\theta_f$, where
$0 <\theta_f< \theta_c$. Then the fat end ($\theta=0$) is no longer a stable
point. If the corresponding bodies are spun with $\theta_{\rm
initial}<\theta_c$, $\theta$ moves to
$\theta_f$ and not to 0, and the bodies will spin at a new stable point
at $\theta_f$. The lemniscate of Bernoulli is not an oval curve, but the
body described by this curve also has a fixed point.

It would be very interesting to make axisymmetric bodies whose cross
sections are described by the Cassini oval (\ref{EqCassiniOval}),
Wassenaar egg curve (\ref{EqWassenaar}) and the lemniscate of Bernoulli
(\ref{EqLemniscate}) and determine if those bodies will spin at a new
stable point that is different from the fat end.

The inclusion of an air chamber at the fat end tends to diminish the
appearance of the fixed point at $\theta_f$. The case of a Cassini oval is
an example. For the Wassenaar egg curve, we need a rather large
air chamber; an empty space for $-a\le z\le -0.5a$ in
Eq.~(\ref{EqWassenaar}) is necessary for the disappearance of $\theta_f$.
The fixed point of the lemniscate vanishes if we take $z_{\min}=-1.33a$ for
Eq.~(\ref{EqLemniscate}). Moreover, the inclusion of an air chamber at the
fat end moves the position of the center of mass $z_g$ toward the thin end
and makes the ratio $A/C$ smaller. Consequently, the values of the
critical angle $\theta_c$ and the inclination angle at rest 
$\theta_r$ move toward $\pi$.

We have assumed Coulomb's law for the friction $F$ (see
Eq.~(\ref{Friction})). If we instead assume a viscous friction law, 
\begin{equation}
F=-{\tilde \mu} MgV_P, \label{ViscousFriction}
\end{equation}
all our conclusions remain unchanged, in particular, the
positions of the critical points and fixed points. Only the
transition time from the unstable to the stable state will be modified.
As an example, the transition time from the angle
$\theta_c$ to $\pi$ is numerically calculated to be 
$t(\theta_c \rightarrow \pi)=(J/\mu M g a^2) \chi$ with a numerical factor 
$\chi\sim {\cal O}(1)$ for Coulomb friction, and $t(\theta_c
\rightarrow
\pi)=(1/{\tilde \mu} g)
\chi$ with $\chi\sim {\cal O}(10)$ for viscous friction .

\begin{acknowledgments}
The author thanks Sinsuke Watanabe and Tsuneo Uematsu for valuable
information on the spinning egg and the tippe top and for critical comments. He
is indebted to Tsuneo Uematsu for reading the manuscript and important
suggestions. Thanks also are due to Shingo Ishiwata, Tetsuji Kuramoto, and
Yoshihiro Shimazu for useful discussions, and Takahiro Ueda for assistance with
figures. This paper is dedicated to the memory of Sozaburo Sasaki.
\end{acknowledgments}

\newpage

\newcommand{\lw}[1]{\smash{\lower2.ex\hbox{#1}}}

\begin{table}[h]
\begin{center}
\begin{tabular}{|c||c|c|c|c|c|}
\hline
\lw{Oval curves} & \multicolumn{1}{c|}{Critical} &\multicolumn{1}{c|}{The
angle} &
\multicolumn{1}{c|}{Fixed point} &\multicolumn{1}{c|}{ Spin at } &
\multicolumn{1}{c|}{ Spin at } 
\\ &\multicolumn{1}{c|}{angle $\theta_c$} &\multicolumn{1}{c|}{at rest
$\theta_r$} &\multicolumn{1}{c|}{angle $\theta_f$}
&\multicolumn{1}{c|}{the fat end} &\multicolumn{1}{c|}{the thin end}\\
\hline\hline Cartesian oval [Eq.~(\ref{EqCartesianOval})]&1.86 & 1.72 & Not
exist & Yes & Yes
\\
\hline Cassini oval [Eq.~(\ref{EqCassiniOval})]& 1.92 & 1.77 & 0.45 & No & Yes \\
\hline
\multicolumn{1}{|c||}{Cassini oval [Eq.~(\ref{EqCassiniOval})] with} &
\lw{1.97} & \lw{1.93} &
\lw{none}& \lw{Yes} & \lw{Yes} \\
\multicolumn{1}{|c||}{an air chamber ($z_{\min}=-1.2a$)} & & & & & \\
\hline Wassenaar egg curve [Eq.~(\ref{EqWassenaar})]& 1.85 & 1.78 & 0.93 & No &
Yes\\
\hline Lemniscate [Eq.~(\ref{EqLemniscate})]& 1.96 & 1.81 & 0.74 & No & No\\
\hline
\end{tabular}
\vspace{0.5cm}
\caption{\label{Summary}Predicted values of the
critical angle
$\theta_c$, the inclination angle at rest $\theta_r$, and the fixed point
angle $\theta_f$ for the axisymmetric bodies whose cross sections are
described by the various oral curves. Also tabulated are the predicted
possibilities of those axisymmetric bodies spinning at the fat and/or the
thin end.}
\end{center}
\end{table}

\newpage
\section*{Figures}

\end{document}